# Understanding of the Renormalization Program in a Mathematically Rigorous Framework and an Intrinsic Mass Scale

**Satish D. Joglekar***
**Department of Physics**
**Indian Institute of Technology, Kanpur**
**208016, U.P., India**
**3 Dec 1999**

**Abstract**

**We show there exists a mathematically consistent framework in which the Renormalization Program can be understood in a natural manner. The framework does not require any violations of mathematical rigor usually associated with the Renormalization program. We use the framework of the non-local field theories [these carry a finite mass scale $\Lambda$]and set up a *finite* perturbative program. We show how this program leads to the perturbation series of the *usual* renormalization program [except one difference] if the series is restructured .We further show that the comparison becomes possible if there exists a finite mass scale $\Lambda$ , with certain properties, in the Quantum Field theory [which we take to be the scale present in the nonlocal theory]. We give a way to estimate the scale $\Lambda$. We also show that the finite perturbation program differs from the usual renormalization program by a term; which we propose can also be used to put a bound on $\Lambda$.**

e-mail:sdj@iitk.ac.in

I INTRODUCTION

The presently successful theory of strong, electromagnetic and weak interactions, viz. the standard model (SM) is a Local Quantum Field Theory (LQFT) [1]. A large body of the successful comparison between the Standard model and the experiments is based upon perturbative calculations. Local Quantum Field Theory calculations, when done perturbatively, are generally plagued with divergences and this certainly holds for the SM calculations [2]. The initial successes of the first LQFT viz. Quantum electrodynamics (QED) were based upon the renormalization program initiated by Feynman, Schrodinger, Tomonaga and developed to a general form by Dyson [3]. This program gives an elaborate procedure for dealing with divergences in LQFT. When this procedure is followed order by order in perturbation theory, it was demonstrated that all the divergences in the theory could be absorbed in the definitions of renormalized fields and parameters as related to the unrenormalized ones. These relations were obtained by imposing by hand "renormalization prescriptions" on the 1PI (proper) vertices which amounted to giving by hand (i.e., from experiment) the physical masses and couplings (and other unphysical parameters). Then the renormalized S-matrix was indeed finite in terms of these. This procedure was highly successful for QED and more so for the further development of Standard Model [2]. It also yielded many results based on renormalization group equations and Callan-Symanzik equation [4].

The renormalization procedure, despite several initial misgivings, came to be regarded as an essential established part of LQFT primarily due to the successes of renormalized LQFT in particle physics. However, as any text book discussion shows, the treatment of divergences in perturbation theory is highly suspicious from the point of view of mathematical rigor [See e.g. Ref.2].

Definition of the infinite Feynman integrals involved requires a regularization. A regularization such as Pauli-Villars violates unitarity for any finite cut off [5], which is recovered only as $\Lambda \to \infty$. Further, in a calculation to any <u>finite</u> order of perturbation theory one makes mathematically unjustified truncations. Thus, in a Pauli-Villars regularization, if

$$Z = 1 + a\, g^2\, \ell n\, \Lambda^2 + O(g^4)$$

$Z^{-1}$ is expanded as

$$Z^{-1} = 1 - a\, g^2\, \ell n\, \Lambda^2 + [a\, g^2\, \ell n\, \Lambda^2]^2 + \ldots\ldots$$

And is truncated to

$$Z^{-1} = 1 - a\, g^2\, \ell n\, \Lambda^2 + O(g^4)$$

Both of which are mathematically invalid operations even for finite (but large enough) $\Lambda$. Similar truncations are made in any regularization. Thus one does not have, in the conventional renormalization procedure of LQFT, unitarity and mathematical consistency for any finite (but large enough) $\Lambda$. Further, the relation(s) such as $0 < Z < 1$ for the wavefunction renormalizations (wherever applicable) obtained from LSZ formulation without recourse to perturbation procedure[6] have to be ignored in these procedures as Z turns out to diverge in perturbation theory [6]. Despite these mathematical shortcoming the renormalization program has succeeded exceedingly well. While it is commonly argued that these may be pathologies introduced by the *perturbation* treatment which may not matter in a non-perturbative treatments, we should recall that much of the success of Standard Model is based on perturbative calculations done following the Renormalization program.

Since early days, one has felt that it may be possible to cure the procedure of these shortcomings; but it has not been possible. However, now, nonlocal formulations of field theories (NLQFT) are possible [7,8] in which the theories have a finite scale $\Lambda$ and are finite (with $\Lambda$ finite), unitary and causal for <u>finite</u> $\Lambda$. These allow us to reconsider the issue of divergences in a new light. We find it convenient to use such a formulation as the background for our line of reasoning.

In such formulations, gauge (and other symmetries) can also be preserved, in a generalized (nonlocal) form [8]. They also admit results of renormalization group at finite $\Lambda$. One can look upon these formulations in two possible ways: (i) as a new nonlocal regularization scheme, an augmentation of the available regularization and renormalization procedures or (ii) as theories in which $\Lambda$ having a fixed <u>finite</u> value serves as the underlying (possibly effective) theory itself. This latter view point has been proposed in Ref. 7,8 and has also been extended and followed up in Ref. 9. In these theories, all calculations are (strictly) finite and (finite) renormalization procedure is needed only for organization of calculations to a given order. We wish to work in the context of such a theory with a finite $\Lambda$. We have demonstrated [10] that in such formulations, the relation $0 < Z < 1$ can in fact be implemented literally and nontrivial conclusions can be drawn from it, which would not be possible in the usual formulation of the renormalization procedure.

We shall demonstrate in this work that we can maintain the mathematical consistency of the formulation in the present approach. Our approach does, in fact, explain where and why the usual renormalization procedure works and elaborate on where we expect the two treatments (to a <u>finite</u> order of perturbation) would give rise to different numerical results. We shall show that we can introduce a rigorous way of defining the perturbation theory which [with a finite $\Lambda$] is a finite process. We shall further show [in Sec.II] that the usual perturbation series is simply obtained by a restructuring of the series so obtained from the above rigorous approach. We show that this depends on the existence of a finite scale $\Lambda$ with certain properties. We also point out a difference between the schemes and further argue that this difference can [from available experimental data] be employed to obtain information about $\Lambda$. This is done in section III.

We outline the approach(es) we want to adopt. We suppose that the particle physics theories are in fact described by a nonlocal action of the type proposed in Ref. [8] with a finite parameter $\Lambda$ present in it [for a brief review of the viewpoint please see ref..10 or 9]. Presence of such a parameter can be looked upon in two ways [9,10]; and we discuss our results in the context of both. In the first approach we may regard $1/\Lambda$ as the scale of nonlocality arizing possibly from a fundamental length scale already existing in nature. In this approach, the NLQFT is an exact theory valid to all energies. In the second approach, which is probably more plausible, the nonlocal theory is looked upon an effective field theory valid upto a certain energy scale (dependent on $\Lambda$) and beyond this scale, the theory would have to be replaced by another NLQFT of a more fundamental nature.

**II AN UNDERSTANDING OF THE RENORMALIZATION PROGRAM IN A RIGOROUS FRAMEWORK**

Many of the calculations done in the context of the Standard Model that are compared with experiments are done perturbatively. In perturbative treatments, one evaluates a physical quantity to a given order n in the coupling constant, say $\lambda$. One carries out renormalizations to this order leaving out any higher order terms both in finite parts *as well as divergences*. In doing so, one is actually ignoring terms, which are a priori much larger [in fact, infinity!]compared to the terms kept. We do not generally have a mathematically sound justification for such a procedure. One then compares this result to the experimental result. In doing so, one has the possibility of choosing a variety of (i) regularization schemes (ii) renormalization conditions/schemes. It is understood that while the results obtained within a given regularization, but using different renormalization conditions may differ in the definitions of its renormalized parameters in terms of its bare parameters, the results are supposed to ultimately agree when summed to all orders[11]. When, however, the series for a physical quantity $P(\lambda)$ is truncated to $O(\lambda^n)$ as is necessary in a practical calculation, the differences (which are supposed to be finite and small) are supposed to arise from higher order finite terms only. Such differences have to be ignored and lead to scheme-dependence. We normally find a good agreement with experiment in innumerable cases and we do not consider the intermediate violations of mathematical rigor important.

In this work, we shall show that we can adopt a mathematically rigorous approach to perturbation theory that enables us to understand why and indeed how this procedure works. This approach is more natural, mathematically sound and less mysterious that the conventional exposition, [which at least a new student finds baffling until he learns to accept it!]. The approach suggested does not require large [or in fact infinite] terms ignored. We wish to further suggest that the very fact that the usual procedure works and leads to results agreeable with experiments has in it information available, ignored otherwise. In fact, the point of view adopted here allows one to deduce the existence of a natural scale in a QFT.

Just to illustrate the violation of mathematical rigor in the renormalization process consider a renormalization constant evaluated to $O[\lambda]$:

$$Z = 1 + A\lambda \ln \Lambda^2/m^2 \qquad (2.1)$$

Suppose we need the inverse $Z^{-1}$, we normally expand:

$$= 1 - A\lambda \ln \frac{\Lambda^2}{m^2} + \left(A\lambda \ln \frac{\Lambda^2}{m^2}\right)^2 + \ldots \qquad (2.2)$$

and keep only the terms of the $O[\lambda]$ in the series:

$$Z^{-1} = 1 - A\lambda \ln \Lambda^2/m^2 \qquad (2.3)$$

to this order of the perturbation series. Such a procedure is normally applied in each order of the perturbation series at various stages of calculation. There are two major violations of mathematical rigor:

[1] Despite the fact that $A\lambda \ln \Lambda^2/m^2$ is [for large enough $\Lambda$] larger than 1 the expansion of the form (2.2) is carried out;

[2] Irrespective of the above, the terms in (2.2): $[A\lambda \ln \Lambda^2/m^2]^2 + \ldots$ which may be comparable [or much larger] to those kept [even though *formally* of higher order] are ignored.

No justification of the above steps in the renormalization program has been given except that the renormalization program so formulated leads to many experimentally verifiable results.

Our aim in this work is to show that the problems posed by the violation of mathematical rigor are avoidable provided that:

[a] A finite scale $\Lambda$ exists with certain properties described later;

[b] The usual perturbation series is understood as a rearrangement of what we would naturally mean by perturbation series which [with a finite $\Lambda$] would be entirely *finite* process and allows a natural formulation [modulo usual ambiguities associated with renormalization conventions].

To formulate this viewpoint, it is in fact convenient to do so in the setting of NLQFT's. To be precise we shall adopt the interpretation of NLQFT's given in the introduction where we regard the scale $\Lambda$ as a finite scale present in the theory either on account of (i) a natural space-time parameter $1/\Lambda$ or (ii) a scale $\Lambda$ characteristic of the range of validity of the theory.

In this viewpoint regarding renormalization, we regard $\Lambda$ as finite and expect the finite renormalizations be carried out rigorously. We do not need to perform mathematical operations that are not rigorous.

We define our procedure for the $n^{th}$ order perturbation theory which is, in fact, what one would do in any finite perturbation scheme and point out the essential differences with the conventional approach. We shall formulate our scheme with reference to QED:

(1) We evaluate a given proper vertex $\Gamma^{(2f,,p)}$ upto n loop approximation. We do calculations directly in terms of the Lagrangian expressed in terms of the unrenormalized parameters. Our results for $\Gamma$ are also expressed in terms of the unrenormalized parameters. For the self-energies and the electron-photon vertex, we determine $Z_1, Z_2, Z_3$ and $\delta m$ by requiring that $Z_2^f Z_3^{p/2} \Gamma^{(2f,,p)}$ satisfies the renormalization conditions upto $n^{th}$ order. We then know that, $Z_2^f Z_3^{p/2} \Gamma^{(2f,,p)}$ gives the correct numerical value of renormalized proper vertex, expressed, however, in terms of unrenormalized quantities.[This is most easily seen with the help of the generating functionals for proper vertices, unrenormalized and renormalized both].

(2) The renormalization conditions give us relations between bare and renormalized quantities. These equations are solved without regarding $\alpha \ln \frac{\Lambda^2}{m^2}$ as a small quantity e.g. for example we do *not* follow the usual steps such as those outlined between (2.1)--(2.3).

(3) We then express $Z_2^f Z_3^{p/2} \Gamma^{(2f,,p)}$ in terms of the renormalized parameters. Usual renormalization theory tells us that if $Z_2^f Z_3^{p/2} \Gamma^{(2f,,p)}$ is expanded by [using a procedure that involves the usage of approximations such as those between (2.1)-(2.3)] upto an order $e^{2\left(\frac{E}{2}-1+n\right)}$ and *the higher order terms neglected*, it will have finite limit as $\Lambda \to \infty$. We, however, required in the present formulation that we <u>do not</u> ignore $O\left(\alpha^{\frac{E}{2}+n}\left(\ln\frac{\Lambda^2}{m^2}\right)^p\right) terms$ and higher as these may be substantial. [The example below will illustrate the differences.]

(4) This procedure may be followed to any desired degree of finite order n.

We make several remarks:

(a) There is no ambiguity on what we mean by $n^{th}$ order perturbation result [ modulo the renormalization convention ambiguities.].

(b) There is no mathematically unjustifiable procedure used or required.

(c) The results upto $n^{th}$ order perturbation theory for this procedure and the standard procedure may differ by terms involving powers of $\alpha \ln \frac{\Lambda^2}{m^2}$ which may be numerically significant.

(d) We shall, however, show that a procedure exists for dealing with the perturbation series that correspond to a restructuring of terms in the perturbation theory as defined above; and moreover it is in fact numerically more accurate way of evaluating the approximation to the quantity under consideration. This procedure gives results which, in fact, *nearly* [see the point (f) below ] correspond to those of the usual interpretation of perturbation theory. Thus the usual renormalization procedure is then understood as a restructuring of the terms in the perturbation series arising from the above rigorous formulation of perturbation theory.

(e) The usual perturbation theory is understood in the above manner provided the theory contains an intrinsic finite mass scale $\Lambda$ such that the above expansions in terms of the coupling constant are rigorously possible.[Please see the discussion in the next section].We emphasize that in our approach, the n-loop perturbation result for the S-matrix is defined irrespective of whether we can carry out the expansion just mentioned. It is only when we want to compare it with the usual perturbation theory rigorously that the need for expansion arises.

(f) There is still, however, left a disagreement with the usual perturbation theory. This disagreement , though it may be small, may yet provide [further] information about the possible intrinsic mass scale present in the theory when compared with experiments. [Please see the discussion in the next section ].This is in addition to the information available from $0 < Z < 1$ [see ref.10 ]in QFT's that may be deduced where this is possible.

(g) With this interpretation in (d), the information obtained about the scale $\Lambda$ may be compatible with that obtained from $0 < Z < 1$ [10] .

It is best to illustrate this point of view with the help of a simple example. Consider QED. We consider the evaluation of one loop correction to e. Let $e_o$ be the bare coupling.

For $\Gamma_\mu$ we shall find

$$\Gamma_\mu( p,p';\Lambda) = \gamma_\mu \{ e_0 + e_0^3 [A \ln \Lambda^2/m^2 + B]\} + e_0^3 f_\mu ( p,p';\Lambda) \qquad (2.4)$$

where we assume that $f_\mu(p,p';\Lambda) = 0$ for $p = p'$ and electrons on mass-shell and is moreover finite if we were to let $\Lambda \to \infty$. We normally define renormalized coupling by the convention [ e = the observed electric charge].

$$\Gamma_\mu(p,p';\Lambda) = e\gamma_\mu \quad \text{at } p-p' = 0 \text{ with } p^2 = p'^2 = m^2 \tag{2.5}$$

[though this is not the only one, we shall stick to it in connection with this example.]. We then find

$$e = e_o + e_0^3 A\left[\ln\frac{\Lambda^2}{m^2} + B\right]$$

$$\equiv e_o[1 + \alpha C(\frac{e_o}{e})^2] \tag{2.6}$$

The above is a *finite* relation so chosen the certain physical quantities [here the electric charge] agree with the one loop result exactly. We have as solution

$$e_0 = e[S_+ + S_-] \tag{2.7}$$

with $S_{+/-} = \{a/2 +/- [a^3/27 + a^2/4]^{1/2}\}^{1/3} \quad ; a = (\alpha C)^{-1}$

We normally truncate this solution as

$$e_o = e[1 - \alpha C + 0\left(\alpha^2 C^2\right)] \tag{2.8}$$

with the clear assumption that $0\left(\alpha^2 C^2\right)$ terms can be ignored. Now suppose, we insist on evaluating one loop result keeping the entire solution (2.7), then we get

$$\Gamma_\mu(p,p') = \gamma_\mu e + e^3 \{S_+ + S_-\}^3 f_\mu(p,p';\Lambda) \tag{2.9}$$

The usual procedure is to truncate $\Gamma_\mu$ to

$$\Gamma_\mu(p,p') = \gamma_\mu e + e^3 \lim f_\mu(p,p';\Lambda) \tag{2.10}$$

[Here, lim refers to $\Lambda \longrightarrow \infty$]. We note that (2.9) and (2.10) differ by terms of the order of $e\alpha^2 C$ which diverge as $\Lambda \rightarrow \infty$ or could well be larger than the terms kept in the usual result (2.10). In addition, (2.9) and (2.10) point to a <u>different dependence</u> on external momenta and hence have observable effects, which could well be large.

How, then, do we get away with the usual perturbative answer of (2.10) ? We explain it in the following manner. The one loop answer (2.9) can also be expanded if $4\alpha C < 1$ [which we assume to be valid: see point (e) above], then (2.9) reads:

$$= e\gamma_\mu + e[\alpha A_{1\mu}(p,p';\Lambda) + \alpha^2 B_{1\mu}[p,p';\alpha,m,\Lambda]] \tag{2.11}$$

where $A_1$ has a finite limit were $\Lambda \rightarrow \infty$ and $B_1$ is a possibly divergent function as $\Lambda \rightarrow \infty$.

Now, imagine working out the result for $\Gamma_\mu(p,p')$ to the next order according the procedure outlined before. From renormalization theory we already know that the result for $\Gamma_\mu(p,p')$ to this order reads:

$$e^{-1}\Gamma_\mu(p,p';\Lambda) = \gamma_\mu + \alpha A_{1\mu}(p,p';m,\Lambda) + [\alpha^2 A_{2\mu}(p,p',m,\Lambda) + \alpha^3 B_{2\mu}[p,p';\alpha,m,\Lambda]] \tag{2.12}$$

The term $B_{2\mu}$ could well be divergent [ or large if $\Lambda$ is large but finite], yet they are normally neglected. Thus, *it is an essential consequence of the results on renormalization* [12] that the application of the perturbation procedure to a further order [second] leads to a contribution to $e^{-1}\Gamma_\mu(p,p';\Lambda)$ of the form:

$$-- \alpha^2 B_{1\mu}[p,p';\alpha,m,;\Lambda] + [\alpha^2 A_{2\mu}(p,p',m;\Lambda) + \alpha^3 B_{2\mu}[p,p';\alpha,m,\Lambda]] \tag{2.13}$$

the first term simply canceling the divergent [ dominant ] term in (2.11). Here, $A_{2\mu}$ has a finite limit as $\Lambda \rightarrow \infty$ and $B_{2\mu}$ may diverge as $\Lambda \rightarrow \infty$. We remark that in (2.13) are contributions that came from the 2-loop diagrams *as well as* the terms arising from the lower order terms from further redefinitions of parameters and the fields.

In a similar manner the next order term cancels the order $\alpha^3$ divergent part of $\alpha^3 B_{2\mu}[p,p';\alpha,m,\Lambda]$; and so on.

We thus see the perturbative expansion procedure, that we outlined, rigorously followed to N loop order, leads to an expansion for $e^{-1}\Gamma_\mu(p,p',\Lambda)$ which reads:

$$e^{-1}\Gamma_\mu(p,p';\Lambda) = \gamma_\mu + \sum_1^N \{-\alpha^n B_{n-1\mu}(p,p';\alpha,m,\Lambda) + [\alpha^n A_{n\mu}(p,p',m;\Lambda) + \alpha^{n+1} B_{n\mu}[p,p';\alpha,m,\Lambda]]\} \quad (2.14)$$

with $B_0 = 0$. We note that the above is only possible if a finite scale $\Lambda$ that allows the series expansions to exist. If we, now, reorder the terms in the series so that the $B_n$ terms from the successive terms are grouped together [which then cancel] we would obtain;

$$e^{-1}\Gamma_\mu(p,p';\Lambda) = \gamma_\mu + \sum_1^N \alpha^n A_{n\mu}(p,p';m,\Lambda) + \alpha^{N+1} B_{N+1\,\mu}[p,p';\alpha,m,\Lambda] \quad (2.15)$$

We now see that the usual perturbative expansion upto $N^{th}$ order

$$e^{-1}\Gamma_\mu(p,p';\Lambda) = \gamma_\mu + \sum_1^N \alpha^n \lim A_{n\mu}(p,p';m,\Lambda)$$

is simply a rearrangement of (2.14), except for the $B_{N+1}$ term and for the limit $\Lambda \longrightarrow \infty$ taken in $A_n$'s.

Evidently, for finite $\Lambda$, even though the $\alpha^{n+1} B_{n\,\mu}[p,p';\alpha,m,\Lambda]$ term could be significant or even dominant as far as the $n^{th}$ order perturbation theory is concerned, they cancel out when an opposite contribution from the next order is taken into account! Thus, the series obtained via the interpretation of the perturbation series outlined earlier, though rigorous, leads to a less convergent series from large oscillating terms; while the usual interpretation of renormalization procedure is simply a reorganization of the same series that converges rapidly and therefore leads to more accurate numerical estimate of the quantity under consideration.

### III: BONDS FOR $\Lambda$ :EXAMPLES OF QED AND QCD

We have seen that the agreement between the usual perturbation theory and the rigorous approach suggested in Sec II depends on the possibility of expansion of the products of renormalization constants and their inverses in powers of the coupling constant which makes the comparison possible in the first place. For example, we can invert some renormalization constant of the form

$$Z = 1+ A\lambda \ln \Lambda^2/m^2 \qquad (3.1)$$

[which for example would be needed in evaluation of the 4-point S-matrix amplitude to 2-loop order] we would need

$$|A\lambda \ln \Lambda^2/m^2| < 1 \qquad (3.2)$$

This implies immediately a bound on how large $\Lambda$ can be. [We have investigated a similar criterion in ref.10 from a different angle.]While we have not investigated such constraints to higher orders for various renormalization constants, we expect a similar bound coming from higher order renormalization constants. From(3.2) we know that we obtain a bound of the " form "[10]

$$\frac{\lambda}{16\pi^2} \ln \frac{\Lambda^2_{max}}{m^2} < 1 \qquad (3.3)$$

This yields,(without worrying about exact coefficients in (3.2) )

$$\Lambda_{max} = m \exp \{ 8\pi^2/\lambda\} \qquad (3.4)$$

For m = 1 GeV, and $\lambda/16\pi^2 = 0.05$ [0.01] we obtain:

$$\Lambda_{max} = 22 \text{ TeV } [10^{18} \text{ TeV}]. \qquad (3.5)$$

Of course, the actual numbers are sensitive to the coefficient in (3.2) and to the value of $\lambda$ in a given theory; however we may expect abound that is testable in near future.

Another possible bound comes from the difference between the usual perturbation theory and the finite scheme proposed in the present work.[13] To $N^{th}$ order of the perturbation series, it is of the form of the last term in (2.15) [there written in the context of the 3-point proper vertex]. Now the value of this term depends on N. We shall now suggest a way to understand what value for N we should choose.

We know from the number of works that the perturbation series is not a convergent series for any value of the coupling in QFT's[14]. Suppose we assume that the perturbation series is an asymptotic series[14]. For a certain value of coupling constant, then there is an optimal number of terms that needs to be kept in the series that gives the best approximation to the physical quantity under consideration. In a given context, let this number be N. Then we shall always compare the usual perturbation series with the result obtained via the procedure adopted here evaluated to $N^{th}$ order. Thus, we shall assume that the quantity under consideration is given actually by a series of the form:

$$e^{-1}\Gamma_\mu (p,p';\Lambda) = \gamma_\mu + \sum_1^N \alpha^n A_{n\mu}(p,p',m;\Lambda) + \alpha^{N+1} B_{N+1\,\mu} [p,p';\alpha,m,\Lambda]] \qquad (3.6)$$

The actual calculation should be compared with the above series: calculation of higher order terms only diminishes the accuracy[15] for an asymptotic series. Thus, when a calculation to order M < N is made in a conventional way, the following result is obtained:

$$e^{-1}\Gamma_\mu (p,p') = \gamma_\mu + \sum_1^N \alpha^n \lim [A_{n\mu}(p,p';m;\Lambda)] \qquad (3.7)$$

where the lim refers to limit $\Lambda \longrightarrow \infty$.

The difference between the the two series (3.6) and (3.7) arises from the following sources:

[1] The last term $\alpha^{N+1} B_{N+1\mu}[p,p';\alpha,m,\Lambda]]$ ;

[2] the perturbation terms of orders (M+1), ………,N;

[3] The differences between $\lim[A_{n\mu}(p,p';m,\Lambda)]$ and $[A_{n\mu}(p,p';m;\Lambda)]$ for $0 < n < M+1$.

While the relative magnitudes of these three terms are dependent on M,N and $\Lambda/m$, we do note the following: The difference [3] above tends to zero as $\Lambda \longrightarrow \infty$, it is likely to be generally ignorable. The term [2] above is owing to the usual higher order perturbation contributions. The last contribution is likely to behave as $\alpha^{N+1}\{\ln[\Lambda^2/m^2]\}^{N+1}$. In case the last contribution is the dominant one, we can suggest a way of obtaining the bound on $\Lambda$ from experimental data.

In this section, we shall illustrate the point by performing some numerical estimates for $\Lambda_{max}$ in the context of two different examples. Here the *purpose is not so much as to obtain accurate numbers, but to explain the principles involved behind these calculations*. We leave accurate evaluations of these quantities using hard experimental data to another detailed work [16].

We shall consider (i) QED (only) and (ii) QCD successively. Again, the stress in either examples is on *illustration* of how a limits <u>can</u> be obtained rather than on exact experimental numbers.

(i) <u>QED</u>:

Suppose, *for the sake of illustration*, the entire contribution to (g-2) of the muon came entirely from QED. Calculations have been done upto $0(\alpha^4)$. calculations. We recall [17] the theoretical and experimental results:

$a_{QED} = 11658480\ (3) \times 10^{-10}$

$a_{expt} = 11659230\ (84) \times 10^{-10}$

The typical experimental error is of the order of $10^{-8}$. The result for $a_{QED}$ is obtained by following the usual renormalization procedure upto 4 loops. The modifications suggested in Sec III say that an uncertainty of $0\left[\left(\dfrac{\alpha}{4\pi}\right)^{N+1}\left(\ell n\dfrac{\Lambda^2}{m^2}\right)^N\right]$ may be present in these results as such terms may not be ignored in fact. Now, we assume that the experimental results agree with the results obtained from the usual perturbation theory. Then we can assume that the extra term $0\left[\left(\frac{\alpha}{4\pi}\right)^{N+1}\left(\ell n\frac{\Lambda^2}{m^2}\right)^N\right]$ is bounded by the experimental error. Further, we assume that for an asymptotic series with expansion parameter $\frac{\alpha}{4\pi}\ell n\frac{\Lambda^2}{m^2}$ needs N+1 = 5 terms [19] for optimal approximation .Then, (without working about the overall constants) we can write a bound of the form [A= a numerical coefficient not evaluated.][20]

$$A\frac{\alpha}{4\pi}\left(\frac{\alpha}{4\pi}\ell n\frac{\Lambda^2}{m^2}\right)^4 \leq 10^{-8}$$

The result is sensitive to the value of A;which we have not determined. As an example,we have [with m =

$M_\mu$ ] and A = 200, $\Lambda_{max} \approx$ 100 TeV.

Of course, the entire procedure is sketchy and this does not make this number very reliable, but the number exists and can be gotten becomes amply evident. [See reference for a detailed treatment.]

Further in view of the fact that the expansion parameter in renormalization constants $\frac{\alpha}{4\pi}\ell n\frac{\Lambda^2}{m^2} \ll 1$, the mathematical operations we do are indeed justifiable.

(ii) <u>QCD</u>:

Suppose we evaluate a quantity X in QCD to $O(\alpha_s^2)$ at energies of $O(M_z)$ Let us suppose that a quantity is calculated as $X = A(\alpha_s/4\pi)^2$. Now $\alpha(M_z)$ is uncertain to within $\pm$ 0.02 i.e. [18]

$\alpha(M_z) = 0.119 \pm 0.02$

Suppose we assume that the usual perturbation results for the quantity agree with the experiment within error bars. Then we know that the last term in (3.6) must be bounded by the experimental uncertainty [which we take as arising from the uncertainty in $\alpha_s$ to illustrate the point] viz. $\frac{2A}{(4\pi)^2}\alpha_s(\Delta\alpha_s)_{expt}$. We next assume that the modification in the renormalization procedure in Sec II introduces additional terms which are of $O\left[\left(\frac{\alpha_s}{4\pi}\right)^{N+1}\left(\ell n\frac{\Delta^2}{m^2}\right)^N\right]$. Suppose we now

assume that these are within error bars of experimental results, then we will have an equation of the form

$$B\left(\frac{\alpha_s}{4\pi}\right)^{N+1}\left(\ell n\frac{\Lambda^2}{m^2}\right)^N \underset{\sim}{<} 2A\alpha_s(\Delta\alpha_s)_{expt}.[4\pi]^{-2}$$

(B and A are unknown, but calculable constants .The details of the actual relation will depend too much on the actual experimental errors and the details of the perturbative calculation which are beyond the scope of the present work. )Numerical estimates can only be done if we knew the details such as the actual experimental uncertainty in a given process, the value of N and constants such as A and B. For (m ~ 1 GeV); we give some representative numbers [19]:

$$\frac{B}{A}=5 \qquad N=4 \qquad \Lambda_{max} \sim 100 \text{ TeV}$$

$$\frac{B}{A}=2 \qquad N=3 \qquad \Lambda_{max} \sim 10 \text{ TeV}$$

Again, we reiterate the remarks that were made for the QED calculation viz., while we have made simplistic assumptions for illustration purposes, such bounds exist and can be obtained is amply illustrated.

**References**

1. See for example, Ta-Pei Cheng and Ling-Fong Li <u>Gauge theory</u> of <u>Elementary Particle Physics</u> (Oxford:Clarendon) (1984).